\documentclass[sn-mathphys,Numbered]{sn-jnl}

\usepackage{graphicx}%
\usepackage{multirow}%
\usepackage{amsmath,amssymb,amsfonts}%
\usepackage{amsthm}%
\usepackage{mathrsfs}%
\usepackage[title]{appendix}%
\usepackage{xcolor}%
\usepackage{textcomp}%
\usepackage{manyfoot}%
\usepackage{booktabs}%
\usepackage{algorithm}%
\usepackage{algorithmicx}%
\usepackage{algpseudocode}%
\usepackage{listings}%
\usepackage{tabularx}%
\usepackage{float}%

\theoremstyle{thmstyleone}%

\theoremstyle{thmstyletwo}%

\theoremstyle{thmstylethree}%
\newtheorem{hypothesis}{Null Hypothesis}

\begin{document}

\title[Article Title]{Co-Pilot for Health:  Personalized Algorithmic AI Nudging to Improve Health Outcomes}


\author[1]{\fnm{Jodi} \sur{Chiam}}\email{jodi.chiam@cuezen.com}
\equalcont{These authors contributed equally to this work.}

\author*[1]{\fnm{Aloysius} \sur{Lim}}\email{aloysius.lim@cuezen.com}
\equalcont{These authors contributed equally to this work.}

\author[1]{\fnm{Cheryl} \sur{Nott}}\email{cheryl.nott@cuezen.com}
\equalcont{These authors contributed equally to this work.}

\author[1,2]{\fnm{Nicholas} \sur{Mark}}\email{nickmmark@gmail.com}
\equalcont{These authors contributed equally}

\author[1]{\fnm{Sunil} \sur{Shinde}}\email{sunil@cuezen.com}
\equalcont{These authors contributed equally to this work.}

\author[1,3]{\fnm{Ankur} \sur{Teredesai}}\email{ankurt@uw.edu}
\equalcont{These authors contributed equally to this work.}

\affil*[1]{\orgname{CueZen, Inc.}, \orgaddress{\city{Seattle}, \postcode{98101}, \state{WA}, \country{USA}}}

\affil[2]{\orgdiv{Department of Critical Care Medicine}, \orgname{Swedish Medical Center}, \orgaddress{\city{Seattle}, \postcode{98122}, \state{WA}, \country{USA}}}

\affil[3]{\orgdiv{Computer Science \& Systems}, \orgname{University of Washington}, \orgaddress{\city{Tacoma}, \postcode{98402}, \state{Washington}, \country{USA}}}


\abstract{The ability to shape health behaviors of large populations automatically, across wearable types and disease conditions at scale has tremendous potential to improve global health outcomes. We designed and implemented an AI driven platform for \textit{digital algorithmic nudging}, enabled by a Graph-Neural Network (GNN) based Recommendation System, and granular health behavior data from wearable fitness devices. Here we describe the efficacy results of this platform with its capabilities of personalized and contextual nudging to $n=84,764$ individuals over a 12-week period in Singapore. We statistically validated that participants in the target group who received such AI optimized daily nudges increased daily physical activity like step count by 6.17\% ($p = 3.09\times10^{-4}$) and weekly minutes of Moderate to Vigorous Physical Activity (MVPA) by 7.61\% ($p = 1.16\times10^{-2}$), compared to matched participants in control group who did not receive any nudges. Further, such nudges were very well received, with a 13.1\% of nudges sent being opened (open rate), and 11.7\% of the opened nudges rated useful compared to 1.9\% rated as not useful thereby demonstrating significant improvement in population level engagement metrics.}




\maketitle

\section{Introduction}\label{sec1}
Non-communicable diseases including cardiovascular complications and cancer are now the major cause of mortality worldwide~\cite{world2004global}.  Modifiable health behaviors, such as physical inactivity, frequent tobacco use, poor sleep hygiene, and consumption of unhealthy foods, are major risk factors for the underlying causes (e.g. obesity, or diabetes) of these non-communicable diseases. Such risk factors can be mitigated significantly by encouraging individual level healthy behavior change, such as staying physically active ~\cite{bakker_sedentary_2021}, curbing unhealthy eating habits ~\cite{marucci_eating_2022}, and smoking cessation ~\cite{zhu_association_2021}. Yet, implementing sustained behavior change for large populations presents many significant challenges and recently mass-nudging has been used in large populations to encourage healthy behaviors at scale. In 2018, SMS nudges were used to increase adherence to a blood pressure monitoring program for postpartum mothers, reducing hospital readmissions by 88\% \cite{hoppe2020telehealth}. In 2020, SMS-based nudges were used to encourage influenza vaccination, resulting in a 5 percent increase in vaccination across a cohort of over 47,000 people in the U.S. \cite{milkman2021flu}. 

Nudge Theory is a behavioral economics concept that influences behavior through the framing of information and choice architecture \cite{Thaler_Sunstein_2009}. Nudges, which are choices presented to steer individuals towards desired outcomes, have tremendous potential to improve health outcomes, and have received increased attention in recent years \cite{purohit_nudging}. Delivering timely nudges to at-risk individuals at population scale can provide a cost-effective intervention that modifies behavior and dramatically improves global health outcomes. In order to be effective, nudges should be \textit{personalized} to the recipient and \textit{contextual} to their circumstances. Providing a ``correct" nudge to at-risk individuals at an opportune moment presents a unique challenge. \textit{Digital algorithmic nudging} enables a highly personalized, context-aware experience to engage individuals and guide them towards setting and achieving personal health goals. In contrast to mass targeting (e.g. billboards, mass emails), digital nudging affords a much more intimate user experience, where nudges can be delivered to participants on a frequent and regular basis. Digital nudging, thus has the potential to enhance engagement, leading to increased and sustained behavior change.

Digital nudging is enabled by Recommendation Systems (RecSys), which are a type of machine learning that utilizes user and item data to provide suggestions that are most pertinent to a particular user. RecSys are ubiquitous in recommending \textit{personalized} content (e.g., Netflix, Spotify, YouTube), products (e.g., Amazon), advertisements (e.g., Facebook Ads), or even potential partners (e.g., online dating). The application of RecSys to healthcare for behavior modification remains nascent, and has recently become easier to deploy due to the increasing popularity of consumer health devices that collect granular biometric data about vital signs (heart rate, pulse oximetry, blood glucose), physical activity (daily steps, minutes of activity), and other health behaviors (hours of sleep, sleep quality). As users go about their daily lives, the real-time data collected by these devices provide rich behavioral and contextual information for RecSys to personalize digital nudges for them.


We theorized that a RecSys would enable better targeting and outcomes than existing methods of engagement by predicting which nudges would be most persuasive in encouraging healthy behaviors for each participant. To understand the effects of nudging on health behaviors, we leveraged an existing mobile health application in Singapore to encourage healthy habits and lifestyles. Through the application, participants could enroll in a physical activity program that encouraged them to achieve a target of 10,000 steps per day and 150 minutes of moderate-to-vigorous physical activity (MVPA) per week. Participants tracked their daily physical activities using consumer fitness trackers and synced their activity data to the mobile application, where they could earn points for completing tiered daily or weekly activity goals. Points could be accumulated and redeemed for rewards such as food, grocery, or transport vouchers. In the 12-month period from Apr 2022 to Mar 2023, there were over 700,000 active participants enrolled in the physical activity program. Traditionally, mass marketing methods such as roadshows, broadcasts, social media, and other promotions had been used to encourage active participation. While past efforts had increased healthy behaviors (e.g. number of steps taken) and participant engagement, they were neither personalized nor contextual. 

We hypothesized that integrating Algorithmic Nudging into an existing mobile health application in Singapore could meaningfully boost program participation and improve health behaviors. To test this hypothesis, we developed a novel RecSys that delivers algorithmically targeted nudges to encourage healthy behaviors among application users.

\section{Results}\label{sec2}

Personalized digital nudges to encourage physical activity (steps and MVPA) were delivered daily over a 12-week period from April 2023 to June 2023 to a randomly sampled subset of active program participants. A total of 84,764 participants were selected to receive nudges: 7,436 (Group 1) were enrolled only in the physical activity program, while 77,328 (Group 2) were enrolled in both the physical activity program and a nutrition program. For each group, a control cohort of participants who did not receive nudges was selected for comparison. The control cohorts were matched on key demographic and baseline behavior attributes. Table \ref{tab:exp_cohorts} summarizes the attributes of the participants in the No Nudges and Nudges cohorts for both groups. During the 12 weeks, all participants continued to receive the existing mass marketing messages sent by the program.

\begin{table}
    \centering
    \begin{tabularx}{\textwidth}{|X|X|X|X|X|} \hline 
         Group&  \multicolumn{2}{|>{\hsize=2\hsize}X|}{Group 1: Participants in physical activity program only}&  \multicolumn{2}{|>{\hsize=2\hsize}X|}{Group 2: Participants in both physical activity and nutrition programs} \\ \hline 
 Group n& \multicolumn{2}{|>{\hsize=2\hsize}X|}{14,901}& \multicolumn{2}{|>{\hsize=2\hsize}X|}{154,766} \\ \hline 
         Treatment&  No Nudges&  Nudges&  No Nudges&  Nudges\\ \hline 
         n&  7,465&  7,436&  77,438&  77,328\\ \hline 
         Mean age&  47.2&  47.6&  46.0&  46.0\\ \hline 
         \% Female&  48.0&  47.7&  62.9&  63.1\\ \hline 
         \% Male&  52.0&  52.3&  37.1&  36.9\\ \hline 
         Mean Daily Steps (start of experiment)&  3,138&  3,153&  4,154&  4,159\\ \hline 
         Mean Weekly MVPA (start of experiment)&  39.5&  40.0&  85.8&  86.8\\ \hline 
         \% iOS&  44.6&  44.5&  43.1&  43.2\\ \hline
         \% Android& 55.4& 55.5& 56.9& 56.8\\ \hline
         \% Apple Watch& 17.2& 16.1& 16.3& 16.5\\ \hline
         \% Fitbit& 6.2& 6.3& 9.8& 9.9\\ \hline
         \% Garmin& 8.0& 7.2& 6.3& 6.1\\ \hline
         \% Samsung Watch& 7.2& 6.9& 9.4& 9.3\\ \hline
         \% Program-issued Fitness Tracker& 61.0& 63.2& 56.8& 56.7\\ \hline
         \% Other Fitness Trackers& 0.4& 0.3& 1.4& 1.5\\ \hline
    \end{tabularx}
    \caption{Participants in control and treatment groups, and their characteristics.}
    \label{tab:exp_cohorts}
\end{table}

Participants' steps and MVPA were recorded using fitness trackers and synced via the mobile health application. The average daily steps, $steps_i^w$ and total weekly minutes of MVPA, $mvpa_i^w$ were computed for each participant $i$, in each week, $0 \leq w \leq 12$. Week 0 is the week immediately prior to the first nudge being sent, and serves as a measurement of participants' baseline behavior before nudging. The following null hypotheses were tested. One-tailed independent-samples t-tests were performed for each group and week. The null hypotheses were rejected if the respective t-tests yielded $p \leq 0.05$.

\begin{hypothesis}
Digital personalized nudges do not have any effect on participants' daily steps:
$\mu_{steps}^{nudge} - \mu_{steps}^{no\_nudge} \leq 0$
\end{hypothesis}

\begin{hypothesis}
Digital personalized nudges do not have any effect on participants' weekly minutes of MVPA:
$\mu_{mvpa}^{nudge} - \mu_{mvpa}^{no\_nudge} \leq 0$
\end{hypothesis}

\begin{figure}
    \centering
    \includegraphics[width=\linewidth]{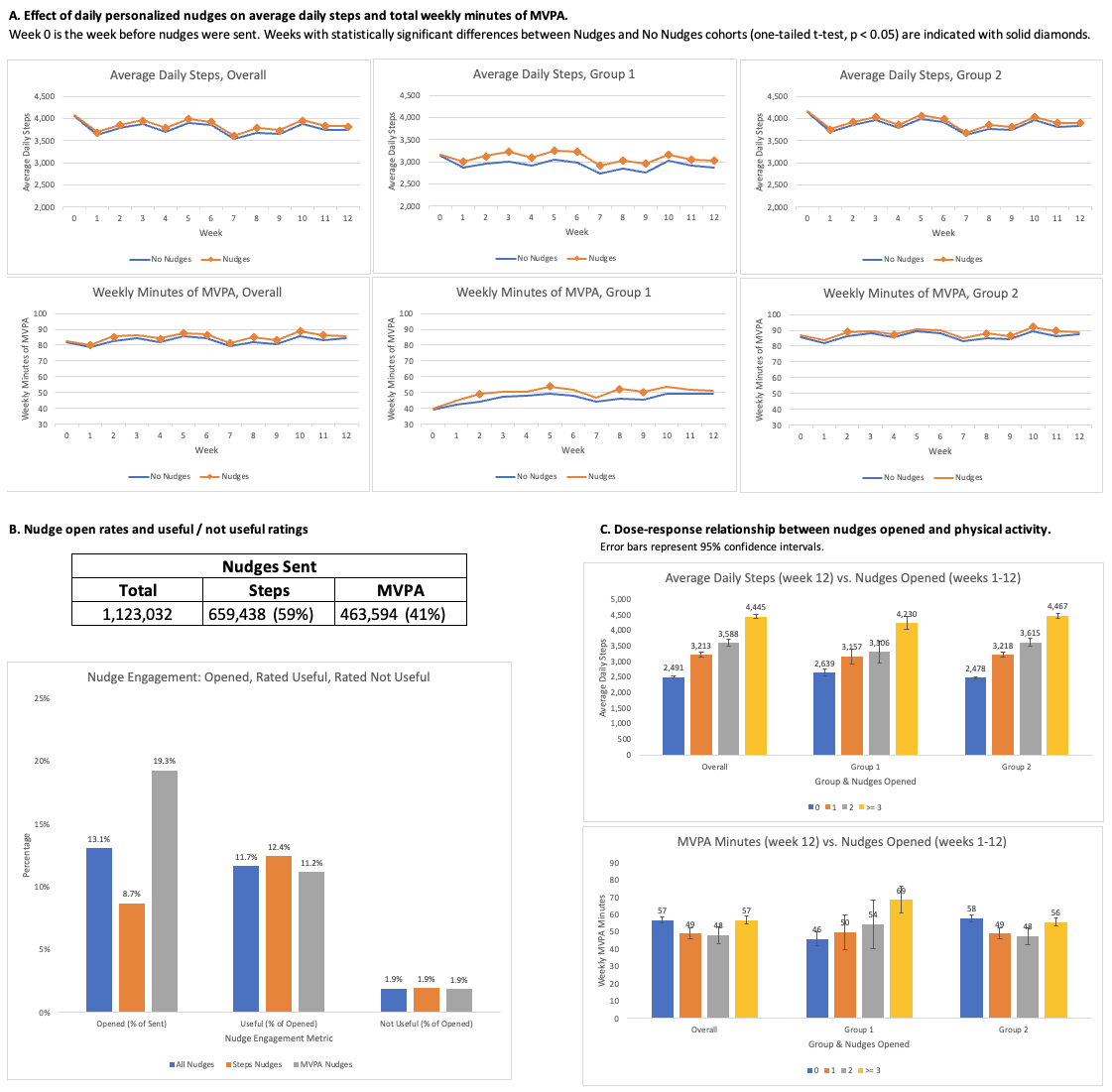}
    \caption{Effect of daily personalized nudges on steps and MVPA, nudge open rates and ratings, and dose-response relationship between nudges opened and physical activity. 
    }
    \label{fig:results}
\end{figure}

\subsection{Effect of Nudges on Physical Activity Behavior}

Figure \ref{fig:results}A shows the physical activity of participants in all groups over the 12 weeks of nudging. Detailed data can be found in Appendix \ref{extended_data}: Extended Data.

Over the entire 12-week period of the study, participants who received nudges did 2.2\% more daily steps (3,829 vs. 3,747 steps; $p = 1.68\times10^{-5}$) and 2.7\% more minutes of MVPA per week (85.0 vs. 82.8 minutes; $p = 9.07\times10^{-3}$) than those who did not get personalized nudges. Therefore, we can reject both null hypotheses that personalized nudges do not impact steps and MVPA behaviors. The effects were larger for Group 1, who were enrolled only in the physical activity program: 6.17\% more daily steps (3,088 vs. 2,909 steps; $p = 3.09\times10^{-4})$ and 7.61\% more weekly minutes of MVPA (50.6 vs. 47.0 minutes; $p = 1.16\times10^{-2})$ for Group 1 participants who received nudges versus those who did not.

The increased physical activity after nudging was persistent across the weeks. Comparing the physical activity of the Nudges cohort versus the No Nudges cohort week by week, participants who received nudges showed a statistically significant increase ($3.76\times10^{-6} \leq p \leq 1.60\times10^{-3}$) in average daily steps for all 12 weeks of the study, suggesting an immediate and sustained effect on behavior. The largest difference was observed at week 8, with a 2.85\% difference (3,783 vs. 3,678 steps; $p = 3.76\times10^{-6}$) between the Nudge and No Nudge cohorts across all participants. The weekly differences were more pronounced for Group 1 participants, reaching 8.37\% (3,226 vs. 2,977 steps; $p = 5.19\times10^{-9}$) at week 6.

There was also an increase in weekly MVPA minutes between the Nudges and No Nudges cohorts in all 12 weeks. These differences were statistically significant for most of the 12 weeks ($1.97\times10^{-3} \leq p \leq 4.63\times10^{-2}$), except for weeks 3 and 12. This suggests that digital nudging can help to increase MVPA in the population for the most part, but it is slightly more difficult to achieve a consistent change of behavior over multiple weeks as compared to steps.

Group 2 participants were inherently more physically active than Group 1 participants, having started at week 0 (prior to nudging) with higher baseline levels of both steps (4,157 vs. 3,146) and MVPA minutes (86.3 vs. 39.7). While both groups increased steps and MVPA minutes with nudging, Group 1 showed larger percentage and absolute differences than Group 2. For example, at week 8, participants in Group 1 who received nudges did 5.95\% or 170 more steps and 12.76\% or 5.92 more minutes of MVPA than those who did not receive nudges, while the differences for Group 2 were 2.62\% or 99 steps and 3.28\% or 2.80 minutes of MVPA respectively. Still, in absolute terms, Group 2 participants did more steps and MVPA than Group 1 participants throughout the 12 weeks.

\subsection{Nudge Engagement}

Over the 12-week period, 1.12 million nudges were sent, comprising approximately 659,000 nudges written to encourage walking more steps, and 464,000 nudges to encourage MVPA (Figure \ref{fig:results}B). The overall open rate was 13.1\%, with 147,000 nudges being opened. MVPA nudges were opened more (19.3\%) than steps nudges (8.7\%). Amongst the nudges opened, they were rated useful 6 times more often than not useful: 11.7\% (17,111) were rated useful, while only 1.9\% (2,790) were rated not useful. This suggests that participants were generally receptive to the nudges that they received.

We also observed a dose response (Figure \ref{fig:results}C), wherein participants who opened more nudges during the 12 weeks walked more steps daily at week 12 -- 0 nudges opened: 2,491 steps; 1 nudge opened: 3,213 steps; 2 nudges opened: 3,588 steps; $\geq$ 3 nudges opened: 4,445 steps. For MVPA, this effect was only seen in Group 1 participants, with 0 nudges opened: 46 minutes; 1 nudge opened: 50 minutes; 2 nudges opened: 54 minutes; $\geq$ 3 nudges opened: 69 minutes.

\section{Discussion}\label{sec3}

We demonstrated the feasibility of using a novel RecSys to deliver over 1.1 million targeted behavioral nudges to a cohort of \textgreater 84,000 people. Compared to un-nudged controls, participants who received nudges took more daily steps and engaged in more minutes of MVPA. Furthermore, the benefits of nudging were sustained across intervention periods and were ``dose-dependent": participants who opened more nudges had greater increases in daily physical activity. The effects were more persistent for steps than for MVPA.

The positive effects of nudging were observed in both participants who started at lower levels of physical activity (Group 1) and those who were already more active at the beginning (Group 2). This indicates that nudging can be a useful mechanism to encourage participants of different starting points to increase and maintain their levels of physical activity. Participants who were initially less active had a greater potential to increase their physical activity to a greater extent, and the results of this study corroborates this, with Group 1 participants increasing in both steps and MVPA to a larger magnitude than Group 2 participants.

The increase in physical activity observed in our cohort with algorithmic targeted nudging was statistically, and likely clinically significant as the effect size could be expected to have salubrious effects based on prior studies. A meta-analysis of 15 studies performed across multiple countries found that cardiovascular mortality begins to decrease when people walk more than 2,300 steps per day \cite{paluch2022daily}. Another meta-analysis found that all-cause mortality begins to decrease after 4,000 steps per day \cite{hupin2015even}.


Increasing MVPA has similar benefits. A UK study using accelerometers \cite{watts2023association} found that a 20 minutes per day increase in MVPA was associated with a decreased risk of hospitalization for several conditions including stroke, diabetes, and pneumonia. A US study found that increasing MVPA by just 30 minutes per week was associated with a 31 percent decrease in the risk of cancer mortality \cite{gilchrist2020association}. Thus, the relatively modest increase in daily steps and weekly MVPA as seen in our study could be expected to have significant population benefits when applied at scale.


The strengths of our study include its scale, integration with an existing mobile health application, and evaluation across multiple participant cohorts. A limitation was that participants had to open the mobile app on their phones to upload tracker data. Thus participants who did not open the app had partial censoring of activity data. This study was also limited to a 12-week period, and longer studies are needed to understand whether digital nudging leads to long-term habituation of behaviors, and whether the behavior change is sustained even if nudges are stopped. Additional longer term studies are needed to determine if the changes in health behavior achieved with algorithmic nudging lead to measurable improvements in health outcomes, such as reductions in cardiovascular events, obesity, and mortality. 

\section{Methods}\label{sec4}

This study aimed to assess the impact of digital personalized nudges on improving health behaviors and participant engagement. It leveraged an existing mobile health application in Singapore in 2023 in a live production setting. Nudges were delivered to a randomly selected subset of active program participants, as part of a 12-week public beta of algorithmic digital nudging.

\subsection{Algorithmic AI Nudging System}

An AI nudging system called NudgeRank\texttrademark{} was developed to automatically collect up-to-date behavior data and generate personalized nudges every day, at scale. Nudges were sent to the mobile health application via push notifications, and subsequent nudge engagement (whether they were opened or rated useful / not useful) and behavior changes were recorded in an automated feedback loop. To isolate the effects of personalized nudges, they were launched in a 12-week public beta to a selected subset of participants. All other aspects of the participants' experience, including the app user experience, program incentives, and exposure to mass marketing messages, remained exactly the same as participants who did not get the personalized nudges.

\begin{figure}
    \centering
    \includegraphics[width=1\linewidth]{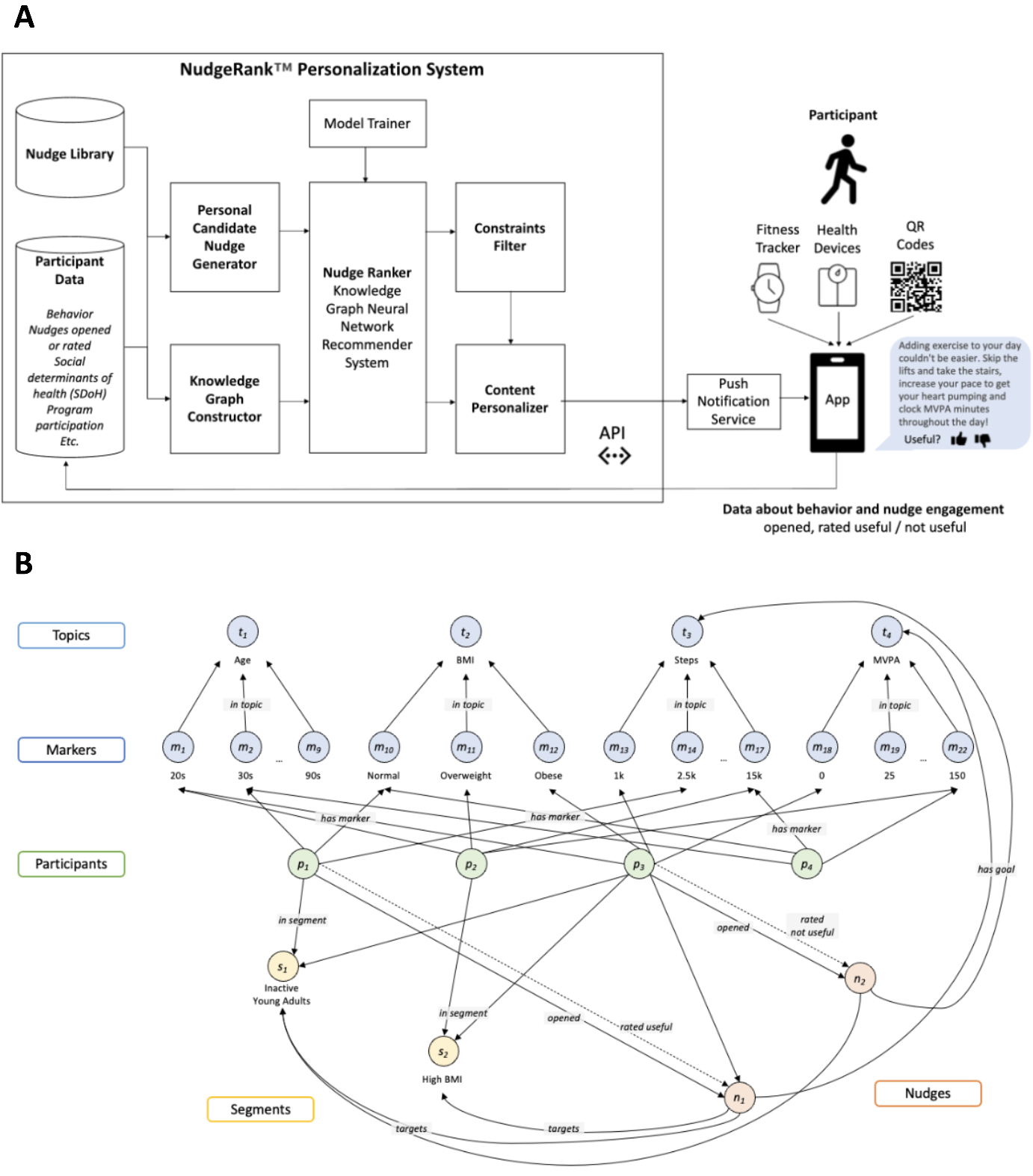}
    \caption{A. Schema of the NudgeRank\texttrademark{} system to deliver personalized algorithmic nudges to participants.
    B. Example of the knowledge graph used to select nudges from the Nudge Library.
    See Methods section for more details.
    }
    \label{fig:nudgerank}
\end{figure}

To generate daily personalized digital nudges at scale, we developed the NudgeRank\texttrademark{} system (Figure \ref{fig:nudgerank}A), composed of several modules operating in tandem. By leveraging distributed computing technologies running on a Kubernetes cluster on commodity virtual servers, the system was designed to easily scale horizontally to generate personalized nudges for millions of participants.

The system first ingests data that are required to generate the nudges. A \textit{Nudge Library} contains the set of nudge templates that have been authored and designed to drive specific behavior changes. Nudge templates can be written with different nudge techniques, tonalities and choice of words, to increase the variety of nudges, since different nudges may appeal to different participants. Nudges in the library can target specific \textit{segments} of participants; for example, a set of nudges can be written specifically for young adults who are overweight and have pre-diabetes. \textit{Participant data} is automatically ingested and updated in the system daily, from a variety of data sources, including daily health behaviors that are synced to the mobile application, demographic or other social determinants of health, health history, and program enrollment and participation data. Data about participants' responses to previous nudges (whether they opened them or rated them with a thumbs up or down) are also automatically tracked and synced from the mobile application. 

The \textit{Personal Candidate Nudge Generator} computes the up-to-date list of candidate nudges for each participant, which adapts to changes in the participant's behavior. Because nudges can be targeted at specific behavioral segments, a participant can receive different nudges as the participant's behavior changes. For example, a set of nudges may be targeted at participants who walk less than 5,000 steps per day. Suppose a participant starts off walking 3,000 steps per day and receives some of these nudges. Two weeks later, the participant increased the daily steps to 8,000; at this point, these nudges are no longer relevant for the participant, but some others may become relevant. By tracking the participant's daily fluctuations in behavior, the system is able to automatically update the dynamic set of candidate nudges for the participant, and ensure that the daily nudges remain contextually relevant for the participant.

The \textit{Knowledge Graph Constructor} processes all the ingested participant and nudge data into a heterogeneous knowledge graph that captures the dynamic interactions between participants, nudges and other knowledge about them. Figure \ref{fig:nudgerank}B illustrates the types of information captured in the knowledge graph. Participants and nudges are represented as nodes in the graph, and directed edges capture the history of participant's past interactions with the nudges. For example, participant \(p_1\) opened nudge \(n_1\) and rated it ``useful"; participant \(p_3\) also opened nudge \(n_1\) but did not give it a rating. Participants' attributes and behaviors are represented by binary \textit{markers}, and the \textit{has marker} edges capture the relationships between participants and their markers. Markers are further classified into \textit{topics} that identify related sets of markers. In the figure, participant  \(p_1\) has the markers \textit{``age: 30s"}, \textit{``BMI: normal"} and \textit{``steps: 2.5k"}. While the figure only illustrates a few markers, the actual graph contains over 130 markers, which can be easily extended as more data sources are ingested. The graph also captures knowledge about the nudges, for example what behavior goal and segments they target. Nudge \(n_2\) targets segment \(s_1\) \textit{``Inactive Young Adults"}, and was written to encourage them to walk more steps. Furthermore, participants \(p_1\) and \(p_3\) are in segment \(s_1\) since they are both in their 20's and 30's, and have not been active in either steps or MVPA. Therefore, nudge \(n_2\) would be a good candidate nudge to send to \(p_1\) and \(p_3\) to encourage them to walk more steps. By capturing all the knowledge about participants and nudges and the relationships between them, the knowledge graph provides a rich representation of the world for the \textit{Nudge Ranker} to finely tailor nudges according to each participant's preferences and behaviors.

The \textit{Nudge Ranker} generates the ranked list of nudges for each participant, from highest to lowest relevance. It is a Knowledge-Graph powered RecSys that uses a Graph Neural Network (GNN) to learn the relationships between users, nudges and their attributes in the graph. It incorporates Knowledge-Aware Attention from Knowledge Graph Attention Network (KGAT) \cite{wang2019kgat}, to consider higher-order relationships between different entities in the graph. For example, the model can consider not only which nudges a participant had previously opened or rated with a thumbs up, but also which nudges other similar participants (with similar markers or segments) had had positive interactions with. The GNN allows these complex inter-relationships to be computed efficiently for a large number of participants and nudges.

We extend existing Knowledge Graph RecSys in two important ways. First, most existing models~\cite{wang2018ripplenet, wang2019knowledge, wang2019kgat, qu2019end} only capture knowledge about items being recommended. We introduce knowledge about users (participants) into the graph to capture information about their behaviors and attributes, which provide rich contextual information about users. Furthermore, our knowledge graph connects users (participants) and items (nudges) not only via their past interactions, but also along other secondary relationships, such as whether a participant belongs to a segment that is also targeted by a nudge. Second, while existing models were designed to work on a static historical graph, our knowledge graph is extremely dynamic and gets updated daily, whenever new data arrives. Edges are added or removed whenever participant behaviors or attributes change; new participants and their edges are added when they download the mobile application for the first time; nudges are added and updated whenever a nudge author edits them in the system. In order to ensure that the recommended nudges remain contextually relevant in such a dynamic data set, an automated \textit{Model Trainer} was developed, comprising machine learning pipelines to train and update the Nudge Ranker's model, so that it can personalize the daily nudges based on the latest version of the knowledge graph.

Once the candidate nudges for each participant have been ranked, the \textit{Constraints Filter} applies several filters to select the final list of nudges to send. Nudge budgets (maximum number of nudges per day) are applied to minimize nudge fatigue. Nudges that have been sent to a participant in the last 7 days are removed, to avoid repetition and increase the diversity of nudges received in a week. Nudges that a participant had rated ``not useful" are removed. All of these constraints are fully configurable according to business requirements.

The final list of nudges for each participant are processed by the \textit{Content Personalizer}, which renders any personalized fields in the nudge templates with the participant's own data. For example, a nudge template \textit{``Great job walking \{\{avg\_daily\_steps\}\} daily steps last week! You are getting pretty close to your goal of 10,000 steps per day!"} can be rendered with each participant's average number of daily steps to make it more personalized.

Finally, the nudges are delivered to the mobile health application via push notifications. The mobile application will track whether and when the nudges are opened or rated (useful or not useful), and collect other data from the participant, such as program data, scans of QR codes, and data from fitness trackers and other smart health devices. All of these data are automatically synced and ingested in the system, as part of a continuous feedback loop.

\subsection{Study Design}

To evaluate the effects of personalized digital nudges on health behaviors, participants were randomly selected from the existing pool of participants who were enrolled in a physical activity program in Singapore. A total of 84,764 existing active participants were selected to receive nudges: 7,436 (Group 1) were enrolled only in the physical activity program,
and 77,328 (Group 2) were also enrolled in a nutrition program. For comparison, control cohorts were selected from the remaining participants to serve as controls; 84,903 participants were selected for the control cohorts, comprising 7,465 control participants for Group 1, and 77,438 for Group 2. Control cohorts were matched on key demographic and baseline behavior data, shown in Table \ref{tab:exp_cohorts}. Statistical tests were performed to ensure that each pair of treatment and control groups were equally matched. Specifically, two-tailed independent samples t-tests were conducted for continuous attributes (age, daily steps at start of experiment, weekly MVPA at start of experiment), and chi-square tests were conducted for categorical attributes (male, female, iOS, Android). All tests returned $p$-values $> 0.05$, indicating that participants in the No Nudges control cohorts were indistinguishable from those in the Nudges cohorts.

In order to assess the impact of nudging on increasing physical activity, all selected participants for the Nudges and No Nudges (control) cohorts must have (a) been at least 18 years old, (b) been recently active in the program, as indicated by at least one sync event of their fitness trackers in the 28 days prior to the start of receiving nudges, and (c) performed less than the recommended 150 minutes of MVPA per week.

All participants continued to experience and engage with the program as-is, including earning points for achieving the steps and MVPA targets, redeeming those points for rewards, and receiving non-personalized messages as part of mass marketing campaigns.

In addition, the Nudges cohort also received one personalized nudge per day that targeted either steps or MVPA, for a period of 12 weeks. A total of 96 nudges were authored by program subject matter experts in the Nudge Library, comprising 31 nudges targeting steps as the behavioral goal and 65 targeting MVPA. The nudges were written to target different levels of achievement as the participants engage in the health behaviors; for example, some nudges remind participants to engage in healthy behaviors if they have been less active recently, while other nudges provide encouragement if they have done well. As participants change in their behaviors during the program, these nudges remain relevant and engaging to them. Nudges were also authored using different nudge techniques such as Framing, Gamification, Reminder and Social Influence. Table \ref{tab:nudges} shows a sample of some of these nudges.

During the 12-week period, data about participants' steps and MVPA behavior were collected via the mobile health application on a daily basis. In addition, nudge engagement (opened nudge, rated useful, rated not useful) data were collected for participants in the Nudge cohort.

The study was reviewed and approved by the relevant authority in Singapore. All participants were existing active participants of their respective programs, and had given prior consent for their fitness data to be collected, for their data to be used to send marketing collateral and personalized messages, nudges or notifications, and for evaluation of results. As this was a quality improvement initiative conducted within an existing mobile health application, additional reviews were not deemed necessary.

\begin{table}
    \centering
    \begin{tabularx}{\textwidth}{|c|c|X|} \hline
         Behavior Goal& Nudge Technique& Nudge\\ \hline

         Steps& Framing& Boost your step count by skipping the lift and taking the stairs when you are out and about! You can move more by adopting this healthy habit.\\ \hline

         Steps& Gamification& Move a little more and aim to hit at least 5,000 steps each day. Get up and take a short walk after every 30 minutes of desk work to gain extra steps.\\ \hline

         Steps& Reminder& Don't let your fitness data go to waste! Sync your fitness tracker to record your progress.\\ \hline

         Steps& Social Influence& 9 in 10 active users clock an average of 5,000 steps daily! Move more and be part of this active community.\\ \hline

         MVPA& Framing& No time to exercise? A quick 15-minute workout only takes up 1\% of your day! Track your progress and stay motivated!\\ \hline

         MVPA& Gamification& Your MVPA minutes in the last seven days is lower than what you've achieved the week before. But don’t throw in the towel! Fire up your motivation, and put in some extra time to work out for the daily prize.\\ \hline

         MVPA& Reminder& Try slotting into your busy schedule a quick 15-minute workout to boost your heart rate.\\ \hline

         MVPA& Social Influence& 2 in 3 active users clock a daily average of at least 10 MVPA minutes! Join them by doing so regularly.\\ \hline

    \end{tabularx}
    \caption{Examples of personalized nudges that were sent to participants.}
    \label{tab:nudges}
\end{table}

\subsection{Code Availability}

Code for the KGAT \cite{wang2019kgat} model used in the recommender system can be found in the original authors' GitHub repository at https://github.com/xiangwang1223/knowledge\_graph\_attention\_network. Other code for the NudgeRank\texttrademark{} system is not publicly available for proprietary reasons.

\section{Data Availability}\label{sec5}

All summary data generated or analyzed during this study are included in this published article and its supplementary information files. Data of individual participants collected or analyzed during this study are not publicly available due to privacy.

\backmatter

\section*{Declarations}

\subsection{Competing Interests}

Authors JC, CN, AL and SS are employees of CueZen, Inc., and declare no non-financial competing interests. Author NM declares no financial or non-financial competing interests. Author AT holds shares in CueZen, Inc. and declares no non-financial competing interests.

\subsection{Author Contributions}

Authors JC, AL, and AT designed the underlying algorithms and AI solution. JC, CN, and AL designed the study, and collected and analyzed the data. Author NM drafted substantive portions of this paper. Authors SS and AT jointly designed the underlying platform and supervised the work. All authors read and approved the final manuscript.

\clearpage

\begin{appendices}

\section{Extended Data}\label{extended_data}

\begin{table}[h!]
    \centering
    \begin{tabular}{|c|c|c|c|c|c|c|c|c|} \hline 
          Week&  \multicolumn{2}{|c|}{No Nudges ($n = 84,903$)}&  \multicolumn{2}{|c|}{Nudges ($n = 84,764$)}&  \multicolumn{2}{|c|}{Difference}& $t$ statistic& $p$-value\\ \hline 
         & Mean&  S.D.&  Mean&  S.D.&  Steps&  \%& & \\ \hline 
          0&  4,064.8&  4,546.6&  4,070.8&  5,344.7&   6.1&  +0.15\%&  0.252&  $4.005\times10^{-1}$ \\ \hline 
          1&  3,626.0&  4,114.6&  3,689.6&  4,752.5&  63.6&  +1.75\%&  2.948&  $\mathbf{1.598\times10^{-3}}$ \\ \hline 
          2&  3,781.9&  4,305.8&  3,856.1&  5,027.5&  74.2&  +1.96\%&  3.265&  $\mathbf{5.469\times10^{-4}}$ \\ \hline 
          3&  3,878.7&  4,303.0&  3,957.7&  5,097.8&  78.9&  +2.03\%&  3.445&  $\mathbf{2.855\times10^{-4}}$ \\ \hline 
          4&  3,701.3&  4,261.4&  3,781.9&  5,016.4&  80.6&  +2.18\%&  3.567&  $\mathbf{1.809\times10^{-4}}$ \\ \hline 
          5&  3,905.5&  4,364.7&  3,993.9&  5,240.0&  88.4&  +2.26\%&  3.774&  $\mathbf{8.028\times10^{-5}}$ \\ \hline 
          6&  3,845.9&  4,360.9&  3,929.8&  5,196.7&  83.9&  +2.18\%&  3.601&  $\mathbf{1.583\times10^{-4}}$ \\ \hline 
          7&  3,548.4&  4,249.2&  3,616.9&  4,599.1&  68.5&  +1.93\%&  3.188&  $\mathbf{7.168\times10^{-4}}$ \\ \hline 
          8&  3,678.2&  4,265.6&  3,783.2&  5,328.4& 105.0&  +2.85\%&  4.479&  $\mathbf{3.756\times10^{-6}}$ \\ \hline 
          9&  3,650.1&  4,296.2&  3,733.4&  5,239.6&  83.2&  +2.28\%&  3.578&  $\mathbf{1.734\times10^{-4}}$ \\ \hline 
         10&  3,871.8&  4,449.5&  3,956.8&  4,985.9&  85.0&  +2.20\%&  3.704&  $\mathbf{1.061\times10^{-4}}$ \\ \hline 
         11&  3,738.6&  4,297.0&  3,829.0&  5,135.2&  90.3&  +2.42\%&  3.929&  $\mathbf{4.263\times10^{-5}}$ \\ \hline 
         12&  3,736.4&  4,245.3&  3,819.7&  5,415.3&  83.3&  +2.23\%&  3.525&  $\mathbf{2.122\times10^{-4}}$ \\ \hline
    \end{tabular}
    \caption{Average daily steps over 12 weeks, for \textbf{all} participants who received nudges ("Nudges" cohort) vs. those who did not receive nudges ("No Nudges" cohort). Week 0 is the week prior to the start of nudging. $p$-values in \textbf{bold} indicate statistical significance ($p \leq 0.05$) in a one-tailed independent samples t-test.}
    \label{tab:steps_overall}
\end{table}

\begin{table}[h!]
    \centering
    \begin{tabular}{|c|c|c|c|c|c|c|c|c|} \hline 
          Week&  \multicolumn{2}{|c|}{No Nudges ($n = 7,465$)}&  \multicolumn{2}{|c|}{Nudges ($n = 7,436$)}&  \multicolumn{2}{|c|}{Difference}& $t$ statistic& $p$-value\\ \hline 
         & Mean&  S.D.&  Mean&  S.D.&  Steps&  \%& & \\ \hline 
          0&  3,138.5&  4,063.0&  3,153.3&  4,064.2&   14.9&  +0.47\%&  0.223&  $4.117\times10^{-1}$ \\ \hline
          1&  2,856.9&  3,727.2&  3,004.0&  3,831.6&  147.1&  +5.15\%&  2.375&  $\mathbf{8.769\times10^{-3}}$ \\ \hline
          2&  2,952.1&  3,809.4&  3,130.0&  3,897.1&  177.9&  +6.03\%&  2.818&  $\mathbf{2.417\times10^{-3}}$ \\ \hline
          3&  3,009.0&  3,777.7&  3,230.5&  3,914.4&  221.5&  +7.36\%&  3.515&  $\mathbf{2.203\times10^{-4}}$ \\ \hline
          4&  2,921.1&  3,792.0&  3,092.6&  4,013.9&  171.5&  +5.87\%&  2.680&  $\mathbf{3.683\times10^{-3}}$ \\ \hline
          5&  3,056.4&  3,953.6&  3,243.1&  4,051.0&  186.7&  +6.11\%&  2.847&  $\mathbf{2.207\times10^{-3}}$ \\ \hline
          6&  2,977.4&  3,791.4&  3,226.4&  4,034.9&  249.1&  +8.37\%&  3.883&  $\mathbf{5.186\times10^{-5}}$ \\ \hline
          7&  2,728.3&  3,646.9&  2,918.3&  3,896.9&  190.1&  +6.97\%&  3.074&  $\mathbf{1.059\times10^{-3}}$ \\ \hline
          8&  2,855.7&  3,719.6&  3,025.5&  3,904.4&  169.8&  +5.95\%&  2.718&  $\mathbf{3.289\times10^{-3}}$ \\ \hline
          9&  2,746.5&  3,708.8&  2,951.5&  3,770.3&  205.0&  +7.47\%&  3.346&  $\mathbf{4.104\times10^{-4}}$ \\ \hline
         10&  3,015.9&  4,049.9&  3,155.1&  4,074.4&  139.2&  +4.62\%&  2.092&  $\mathbf{1.824\times10^{-2}}$ \\ \hline
         11&  2,914.6&  3,815.0&  3,045.6&  3,975.5&  130.9&  +4.49\%&  2.051&  $\mathbf{2.015\times10^{-2}}$ \\ \hline
         12&  2,869.3&  3,823.8&  3,034.9&  3,966.6&  165.6&  +5.77\%&  2.595&  $\mathbf{4.735\times10^{-3}}$ \\ \hline
    \end{tabular}
    \caption{Average daily steps over 12 weeks, for \textbf{Group 1} participants (enrolled in the physical activity program only) who received nudges ("Nudges" cohort) vs. those who did not receive nudges ("No Nudges" cohort). Week 0 is the week prior to the start of nudging. $p$-values in \textbf{bold} indicate statistical significance ($p \leq 0.05$) in a one-tailed independent samples t-test.}
    \label{tab:steps_g1}
\end{table}

\begin{table}[h!]
    \centering
    \begin{tabular}{|c|c|c|c|c|c|c|c|c|} \hline 
          Week&  \multicolumn{2}{|c|}{No Nudges ($n = 77,438$)}&  \multicolumn{2}{|c|}{Nudges ($n = 77,328$)}&  \multicolumn{2}{|c|}{Difference}& $t$ statistic& $p$-value\\ \hline 
         & Mean&  S.D.&  Mean&  S.D.&  Steps&  \%& & \\ \hline 
          0&  4,154.1&  4,580.7&  4,159.1&  5,443.9&   5.0&  +0.12\%&  0.196&  $4.224\times10^{-1}$ \\ \hline
          1&  3,700.1&  4,142.5&  3,755.5&  4,826.7&  55.4&  +1.50\%&  2.424&  $\mathbf{7.679\times10^{-3}}$ \\ \hline
          2&  3,861.9&  4,342.4&  3,925.9&  5,117.7&  64.0&  +1.66\%&  2.654&  $\mathbf{3.976\times10^{-3}}$ \\ \hline
          3&  3,962.6&  4,341.1&  4,027.6&  5,192.1&  65.0&  +1.64\%&  2.671&  $\mathbf{3.780\times10^{-3}}$ \\ \hline
          4&  3,776.5&  4,296.5&  3,848.2&  5,097.6&  71.7&  +1.90\%&  2.991&  $\mathbf{1.392\times10^{-3}}$ \\ \hline
          5&  3,987.3&  4,393.7&  4,066.1&  5,334.9&  78.7&  +1.97\%&  3.168&  $\mathbf{7.672\times10^{-4}}$ \\ \hline
          6&  3,929.6&  4,402.9&  3,997.4&  5,290.1&  67.8&  +1.73\%&  2.740&  $\mathbf{3.073\times10^{-3}}$ \\ \hline
          7&  3,627.5&  4,294.5&  3,684.1&  4,655.6&  56.7&  +1.56\%&  2.488&  $\mathbf{6.422\times10^{-3}}$ \\ \hline
          8&  3,757.5&  4,306.3&  3,856.0&  5,440.2&  98.5&  +2.62\%&  3.950&  $\mathbf{3.905\times10^{-5}}$ \\ \hline
          9&  3,737.2&  4,338.7&  3,808.5&  5,353.7&  71.3&  +1.91\%&  2.878&  $\mathbf{1.999\times10^{-3}}$ \\ \hline
         10&  3,954.3&  4,477.6&  4,033.9&  5,058.3&  79.6&  +2.01\%&  3.277&  $\mathbf{5.256\times10^{-4}}$ \\ \hline
         11&  3,818.1&  4,332.3&  3,904.3&  5,227.0&  86.2&  +2.26\%&  3.533&  $\mathbf{2.053\times10^{-4}}$ \\ \hline
         12&  3,820.0&  4,274.5&  3,895.1&  5,528.9&  75.1&  +1.97\%&  2.991&  $\mathbf{1.390\times10^{-3}}$ \\ \hline
    \end{tabular}
    \caption{Average daily steps over 12 weeks, for \textbf{Group 2} participants (enrolled in both the physical activity and nutrition programs) who received nudges ("Nudges" cohort) vs. those who did not receive nudges ("No Nudges" cohort). Week 0 is the week prior to the start of nudging. $p$-values in \textbf{bold} indicate statistical significance ($p \leq 0.05$) in a one-tailed independent samples t-test.}
    \label{tab:steps_g2}
\end{table}

\begin{table}[h!]
    \centering
    \begin{tabular}{|c|c|c|c|c|c|c|c|c|} \hline 
          Week&  \multicolumn{2}{|c|}{No Nudges ($n = 84,903$)}&  \multicolumn{2}{|c|}{Nudges ($n = 84,764$)}&  \multicolumn{2}{|c|}{Difference}& $t$ statistic& $p$-value\\ \hline 
         & Mean&  S.D.&  Mean&  S.D.&  MVPA Minutes&  \%& & \\ \hline 
          0&  81.76&  218.62&  82.73&  227.71&  0.96&  +1.18\%&  0.888&  $1.874\times10^{-1}$ \\ \hline 
          1&  78.49&  208.15&  80.29&  216.06&  1.79&  +2.29\%&  1.741&  $\mathbf{4.082\times10^{-2}}$ \\ \hline 
          2&  82.31&  216.17&  85.42&  228.36&  3.11&  +3.78\%&  2.884&  $\mathbf{1.965\times10^{-3}}$ \\ \hline 
          3&  84.69&  223.41&  86.07&  226.57&  1.38&  +1.63\%&  1.260&  $1.038\times10^{-1}$ \\ \hline 
          4&  82.11&  219.72&  84.19&  226.55&  2.07&  +2.53\%&  1.914&  $\mathbf{2.783\times10^{-2}}$ \\ \hline 
          5&  85.62&  229.51&  87.61&  231.55&  1.99&  +2.32\%&  1.776&  $\mathbf{3.789\times10^{-2}}$ \\ \hline 
          6&  84.61&  230.30&  86.73&  233.09&  2.11&  +2.50\%&  1.879&  $\mathbf{3.014\times10^{-2}}$ \\ \hline 
          7&  79.51&  226.09&  81.38&  232.98&  1.87&  +2.35\%&  1.682&  $\mathbf{4.628\times10^{-2}}$ \\ \hline 
          8&  81.80&  224.99&  84.87&  239.32&  3.08&  +3.76\%&  2.730&  $\mathbf{3.172\times10^{-3}}$ \\ \hline 
          9&  80.91&  220.90&  83.40&  227.48&  2.49&  +3.07\%&  2.285&  $\mathbf{1.117\times10^{-2}}$ \\ \hline 
         10&  85.74&  235.03&  88.67&  241.26&  2.93&  +3.42\%&  2.533&  $\mathbf{5.648\times10^{-3}}$ \\ \hline 
         11&  83.26&  228.45&  86.35&  242.05&  3.09&  +3.71\%&  2.704&  $\mathbf{3.430\times10^{-3}}$ \\ \hline 
         12&  84.34&  237.44&  85.51&  240.71&  1.16&  +1.38\%&  1.003&  $1.580\times10^{-1}$ \\ \hline
    \end{tabular}
    \caption{Weekly minutes of MVPA over 12 weeks, for \textbf{all} participants who received nudges ("Nudges" cohort) vs. those who did not receive nudges ("No Nudges" cohort). Week 0 is the week prior to the start of nudging. $p$-values in \textbf{bold} indicate statistical significance ($p \leq 0.05$) in a one-tailed independent samples t-test.}
    \label{tab:mvpa_overall}
\end{table}

\begin{table}[h!]
    \centering
    \begin{tabular}{|c|c|c|c|c|c|c|c|c|} \hline 
          Week&  \multicolumn{2}{|c|}{No Nudges ($n = 7,465$)}&  \multicolumn{2}{|c|}{Nudges ($n = 7,436$)}&  \multicolumn{2}{|c|}{Difference}& $t$ statistic& $p$-value\\ \hline 
         & Mean&  S.D.&  Mean&  S.D.&  MVPA Minutes&  \%& & \\ \hline 
          0&  39.53&   79.98&  39.95&   78.37&  0.42&   +1.05\%&  0.321&  $3.740\times10^{-1}$ \\ \hline
          1&  42.65&  108.42&  44.86&  101.77&  2.21&   +5.18\%&  1.281&  $1.000\times10^{-1}$ \\ \hline
          2&  44.25&  109.80&  49.00&  123.38&  4.75&  +10.74\%&  2.484&  $\mathbf{6.499\times10^{-3}}$ \\ \hline
          3&  47.41&  127.02&  50.58&  123.62&  3.17&   +6.69\%&  1.544&  $6.128\times10^{-2}$ \\ \hline
          4&  47.90&  125.28&  50.64&  123.65&  2.74&   +5.71\%&  1.341&  $8.992\times10^{-2}$ \\ \hline
          5&  49.24&  137.77&  53.73&  133.59&  4.49&   +9.12\%&  2.019&  $\mathbf{2.175\times10^{-2}}$ \\ \hline
          6&  48.15&  134.07&  51.70&  132.50&  3.56&   +7.38\%&  1.628&  $5.176\times10^{-2}$ \\ \hline
          7&  44.18&  135.71&  46.82&  141.13&  2.65&   +5.99\%&  1.171&  $1.209\times10^{-1}$ \\ \hline
          8&  46.43&  135.46&  52.36&  166.08&  5.92&  +12.76\%&  2.385&  $\mathbf{8.544\times10^{-3}}$ \\ \hline
          9&  45.49&  129.29&  50.41&  140.63&  4.92&  +10.82\%&  2.224&  $\mathbf{1.307\times10^{-2}}$ \\ \hline
         10&  49.43&  146.95&  53.45&  151.63&  4.02&   +8.14\%&  1.644&  $5.010\times10^{-2}$ \\ \hline
         11&  49.51&  160.19&  51.99&  145.40&  2.48&   +5.01\%&  0.990&  $1.611\times10^{-1}$ \\ \hline
         12&  49.08&  142.48&  51.08&  144.27&  2.00&   +4.07\%&  0.849&  $1.978\times10^{-1}$ \\ \hline
    \end{tabular}
    \caption{Weekly minutes of MVPA over 12 weeks, for \textbf{Group 1} participants (enrolled in the physical activity program only) who received nudges ("Nudges" cohort) vs. those who did not receive nudges ("No Nudges" cohort). Week 0 is the week prior to the start of nudging. $p$-values in \textbf{bold} indicate statistical significance ($p \leq 0.05$) in a one-tailed independent samples t-test.}
    \label{tab:mvpa_g1}
\end{table}

\begin{table}[h!]
    \centering
    \begin{tabular}{|c|c|c|c|c|c|c|c|c|} \hline 
          Week&  \multicolumn{2}{|c|}{No Nudges ($n = 77,438$)}&  \multicolumn{2}{|c|}{Nudges ($n = 77,328$)}&  \multicolumn{2}{|c|}{Difference}& $t$ statistic& $p$-value\\ \hline 
         & Mean&  S.D.&  Mean&  S.D.&  MVPA Minutes&  \%& & \\ \hline 
          0&  85.84&  227.15&  86.84&  236.76&  1.00&  +1.17\%&  0.852&  $1.972\times10^{-1}$ \\ \hline
          1&  81.95&  215.02&  83.69&  223.71&  1.75&  +2.13\%&  1.565&  $5.883\times10^{-2}$ \\ \hline
          2&  85.98&  223.43&  88.92&  235.71&  2.95&  +3.43\%&  2.524&  $\mathbf{5.804\times10^{-3}}$ \\ \hline
          3&  88.28&  230.26&  89.48&  233.81&  1.20&  +1.35\%&  1.014&  $1.554\times10^{-1}$ \\ \hline
          4&  85.41&  226.49&  87.41&  233.82&  2.00&  +2.34\%&  1.711&  $\mathbf{4.357\times10^{-2}}$ \\ \hline
          5&  89.13&  236.18&  90.87&  238.61&  1.74&  +1.95\%&  1.441&  $7.486\times10^{-2}$ \\ \hline
          6&  88.13&  237.23&  90.10&  240.29&  1.97&  +2.23\%&  1.620&  $5.264\times10^{-2}$ \\ \hline
          7&  82.91&  232.68&  84.70&  239.71&  1.79&  +2.16\%&  1.492&  $6.789\times10^{-2}$ \\ \hline
          8&  85.21&  231.52&  88.00&  244.98&  2.80&  +3.28\%&  2.308&  $\mathbf{1.051\times10^{-2}}$ \\ \hline
          9&  84.33&  227.51&  86.57&  233.89&  2.24&  +2.66\%&  1.914&  $\mathbf{2.784\times10^{-2}}$ \\ \hline
         10&  89.24&  241.54&  92.05&  247.91&  2.82&  +3.16\%&  2.263&  $\mathbf{1.181\times10^{-2}}$ \\ \hline
         11&  86.51&  233.72&  89.65&  249.13&  3.14&  +3.63\%&  2.557&  $\mathbf{5.279\times10^{-3}}$ \\ \hline
         12&  87.74&  244.39&  88.82&  247.76&  1.08&  +1.23\%&  0.860&  $1.949\times10^{-1}$ \\ \hline
    \end{tabular}
    \caption{Weekly minutes of MVPA over 12 weeks, for \textbf{Group 2} participants (enrolled in both the physical activity and nutrition programs) who received nudges ("Nudges" cohort) vs. those who did not receive nudges ("No Nudges" cohort). Week 0 is the week prior to the start of nudging. $p$-values in \textbf{bold} indicate statistical significance ($p \leq 0.05$) in a one-tailed independent samples t-test.}
    \label{tab:mvpa_g2}
\end{table}

\begin{table}[h!]
    \centering
    \begin{tabularx}{\textwidth}{|X|X|X|X|X|X|} \hline 
          Nudges Opened (Weeks 1-12)&  $n$&  \multicolumn{2}{|c|}{Average Daily Steps (Weeks 1-12)}&  \multicolumn{2}{|c|}{Total MVPA Minutes (Week 12)}\\ \hline 
         &  &  Mean&  S.D.&  Mean&  S.D. \\ \hline 
                0&  54,562&  2,491.48&  4,296.21&  56.91&  200.96\\ \hline
                1&  10,867&  3,212.84&  3,807.07&  49.22&  167.17\\ \hline
                2&   5,385&  3,587.72&  4,150.51&  48.23&  178.31\\ \hline
         $\geq$ 3&  13,950&  4,444.72&  4,129.64&  57.00&  136.61\\ \hline
    \end{tabularx}
    \caption{Average daily steps and weekly minutes of MVPA in Week 12, versus total number of nudges opened in weeks 1 to 12, for \textbf{all} participants who received nudges.}
    \label{tab:opened_v_pa_all}
\end{table}

\begin{table}[h!]
    \centering
    \begin{tabularx}{\textwidth}{|X|X|X|X|X|X|} \hline 
          Nudges Opened (Weeks 1-12)&  $n$&  \multicolumn{2}{|c|}{Average Daily Steps (Weeks 1-12)}&  \multicolumn{2}{|c|}{Total MVPA Minutes (Week 12)}\\ \hline 
         &  &  Mean&  S.D.&  Mean&  S.D. \\ \hline 
                0&  4,649&  2,639.05&  3,897.80&  45.95&  138.43\\ \hline
                1&    972&  3,156.61&  3,951.75&  49.89&  163.06\\ \hline
                2&    483&  3,305.84&  3,995.83&  54.48&  158.70\\ \hline
         $\geq$ 3&  1,332&  4,229.52&  3,955.43&  68.63&  142.93\\ \hline
    \end{tabularx}
    \caption{Average daily steps and weekly minutes of MVPA in Week 12, versus total number of nudges opened in weeks 1 to 12, for \textbf{Group 1} participants (enrolled in the physical activity program only) who received nudges.}
    \label{tab:opened_v_pa_g1}
\end{table}

\begin{table}[h!]
    \centering
    \begin{tabularx}{\textwidth}{|X|X|X|X|X|X|} \hline 
          Nudges Opened (Weeks 1-12)&  $n$&  \multicolumn{2}{|c|}{Average Daily Steps (Weeks 1-12)}&  \multicolumn{2}{|c|}{Total MVPA Minutes (Week 12)}\\ \hline 
         &  &  Mean&  S.D.&  Mean&  S.D. \\ \hline 
                0&  49,913&  2,477.73&  4,331.23&  57.93&  205.79\\ \hline
                1&   9,895&  3,218.36&  3,792.72&  49.15&  167.58\\ \hline
                2&   4,902&  3,615.49&  4,164.80&  47.61&  180.13\\ \hline
         $\geq$ 3&  12,618&  4,467.44&  4,147.11&  55.77&  135.87\\ \hline
    \end{tabularx}
    \caption{Average daily steps and weekly minutes of MVPA in Week 12, versus total number of nudges opened in weeks 1 to 12, for \textbf{Group 2} participants (enrolled in both the physical activity and nutrition programs) who received nudges.}
    \label{tab:opened_v_pa_g2}
\end{table}

\begin{table}[h!]
    \centering
    \begin{tabularx}{\textwidth}{|c|X|X|X|X|X|X|X|} \hline
        Nudge Technique& Sent& Opened& Opened / Sent& Useful& Useful / Opened& Not Useful& Not Useful / Opened \\ \hline
        Framing           &   333,614&  46,413& 13.9\%&  5,553& 12.0\%&   759& 1.6\% \\ \hline
        Gamification      &   218,428&  28,816& 13.2\%&  3,322& 11.5\%&   564& 2.0\% \\ \hline
        Reminder          &   339,078&  38,459& 11.3\%&  4,443& 11.6\%&   804& 2.1\% \\ \hline
        Social Influence  &   231,912&  33,000& 14.2\%&  3,793& 11.5\%&   663& 2.0\% \\ \hline
        Total             & 1,123,032& 146,688& 13.1\%& 17,111& 11.7\%& 2,790& 1.9\% \\ \hline
    \end{tabularx}
    \caption{Nudge engagement: nudges sent, opened, rated useful or not useful for \textbf{all} nudges and by nudge technique.}
    \label{tab:nudge_engagement_all}
\end{table}

\begin{table}[h!]
    \centering
    \begin{tabularx}{\textwidth}{|c|X|X|X|X|X|X|X|} \hline
        Nudge Technique& Sent& Opened& Opened / Sent& Useful& Useful / Opened& Not Useful& Not Useful / Opened \\ \hline
        Framing           & 157,490& 12,741& 8.1\%& 1,655& 13.0\%&   240& 1.9\% \\ \hline
        Gamification      & 137,652& 12,811& 9.3\%& 1,506& 11.8\%&   226& 1.8\% \\ \hline
        Reminder          & 236,098& 19,092& 8.1\%& 2,332& 12.2\%&   407& 2.1\% \\ \hline
        Social Influence  & 128,198& 12,697& 9.9\%& 1,632& 12.9\%&   245& 1.9\% \\ \hline
        Total             & 659,438& 57,341& 8.7\%& 7,125& 12.4\%& 1,118& 1.9\% \\ \hline
    \end{tabularx}
    \caption{Nudge engagement: nudges sent, opened, rated useful or not useful for \textbf{steps} nudges and by nudge technique.}
    \label{tab:nudge_engagement_steps}
\end{table}

\begin{table}[h!]
    \centering
    \begin{tabularx}{\textwidth}{|c|X|X|X|X|X|X|X|} \hline
        Nudge Technique& Sent& Opened& Opened / Sent& Useful& Useful / Opened& Not Useful& Not Useful / Opened \\ \hline
        Framing           & 176,124& 33,672& 19.1\%& 3,898& 11.6\%&  519& 1.5\% \\ \hline
        Gamification      &  80,776& 16,005& 19.8\%& 1,816& 11.3\%&  338& 2.1\% \\ \hline
        Reminder          & 102,980& 19,367& 18.8\%& 2,111& 10.9\%&  397& 2.0\% \\ \hline
        Social Influence  & 103,714& 20,303& 19.6\%& 2,161& 10.6\%&  418& 2.1\% \\ \hline
        Total             & 463,594& 89,347& 19.3\%& 9,986& 11.2\%& 1,672& 1.9\% \\ \hline
    \end{tabularx}
    \caption{Nudge engagement: nudges sent, opened, rated useful or not useful for \textbf{mvpa} nudges and by nudge technique.}
    \label{tab:nudge_engagement_mvpa}
\end{table}

\end{appendices}


\clearpage

\bibliography{sn-article}

\end{document}